\documentclass[preprint,showpacs,preprintnumbers,amsmath,amssymb]{revtex4}

\begin{document}

\title{Elucidation of Conformational Hysteresis on a Giant DNA}

\author {Chwen-Yang Shew \footnote{Electronic mail: shew@mail.csi.cuny.edu}}

\affiliation{Department of Chemistry and Graduate Center, City University of New York, College of Staten Island, 2800 Victory Boulevard, Staten Island, NY 10314, USA}

\author {Yuji Higuchi and Kenichi Yoshikawa}

\affiliation{Department of Physics, Graduate School of Science, Kyoto University, Kyoto 606-8502, Japan}

\begin{abstract}
The conformational behavior of a giant DNA mediated by condensing agents in the bulk solution has been investigated through experimental and theoretical approaches. Experimentally, a pronounced conformational hysteresis is observed for folding and unfolding processes, by increasing and decreasing the concentration of condensing agent PEG (Polyethylene glycol), respectively. To elucidate the observed hysteresis, a semiflexible chain model is studied by using Monte Carlo simulations for the coil-globule transition. In the simulations, the hysteresis loop emerges for stiff enough chains, indicating distinct pathways for folding and unfolding processes. Also, our results show that globular state is thermodynamically more stable than coiled state in the hysteresis loop. Our findings suggest that increasing chain stiffness may reduce the chain conformations relevant to the folding pathway, which impedes the folding process.  
\end{abstract}



\pacs{64.90.+b, 87.15.He, 87.53.Wz}

\maketitle

\newpage

\section{Introduction}

The conformational behavior of a giant DNA has been actively studied over past years because its unique morphology is relevant to the fundamental physics of life science. For example, the experimental evidence has linked the conformational transition of a long chain DNA with the switching of transcriptional activities on-and-off {\em in vivo} \cite{dna1,rev05,bloom,dna2}, and with the rate of DNA polymerase. \cite{dnapoly} It is known that elongation of a compact DNA is important to facilitate the above cellular activities in the presence of cellular proteins. \cite{dna1,dna2,dnapoly} The cellular proteins play the role on altering the media condition surrounding a DNA molecule. Such a system is similar to the fundamental coil-to-globule conformational transition in polymer physics. \cite{lifshitz} 

Experimentally, the coiled and globular DNA can be induced by changing the amount of condensing agent, for instance high valent cations or neutral PEG (Polyethylene glycol), etc. More than a decade ago, by using the single molecular fluorescence image, it has been found that the coil-to-globule transition of DNA is of the first-order (all-or-none) phase transition with coexistent coiled and globular conformational states \cite{dna1,dnapoly}. The change in the effective volume is of the order $10^4$-$10^5$. \cite{dna1} In the discrete conformational transition, there exists a region of coexistent elongated and compact DNA chains. Since the effective segmental density in the elongated state is very low compared to that in the compact state, the precise intensity on the light scattering is very weak. This means that the physico-chemical properties of the ensemble of chains, such as light scattering of a DNA solution, seem continuous. \cite{dna1} To this end, the single molecular fluorescence image remains the most suitable experimental method for detecting single chain conformations.

When the concentration of condensing agents is not in large excess, the compact DNA chains are soluble and do not aggregate with each other even through frequent collisions in solution. Under this condition, the kinetic processes of nucleation and growth can also be observed \cite{kin1}. Through such processes, a regular toroidal structure is generated as the product of chain compaction. However, when a large excess of condensing agent is added to the DNA solution, a rapid collapse is induced by the formation of a spherical product via irregular packing of the chain. \cite{dna3} Such behavior of single DNA molecules is similar to the kinetic aspects of crystallization from a supercooled solution: when the degree of supercooling is small, a regular pure crystal is obtained; on the other hand, when the degree is large, a ``dirty'' precipitate is obtained (spinodal decomposition). In compact particles with regular tight packing, such as a toroid and rod, the negative charge disappears completely except on the surface \cite{dnacharge}. The surviving negative charge on the surface of the compact DNA enhances its soluble colloid properties.

From the theoretical standpoint, the unique conformational behavior of a giant DNA chain has been mainly attributed to its semiflexibility. \cite{dna1,dna3} The simulation results have revealed that the coil-to-globule transition is continuous for flexible chains, but is discontinuous for stiff chains like DNA. \cite{dna3} Also, unlike the liquid-like globule for a collapsed flexible chain, an ordered (crystal-like) toroid structure can be formed in the compact state of DNA. \cite{dna2,sim1} These findings are consistent with experimental observations. \cite{dna1,bloom,dna3} 

In addition to characterization of the biphasic DNA conformations, the dynamics involved in the conformational transition has also attracted a great deal of attention. The techniques by using extensional force and extensional flow methods have been developed to stretch a DNA chain. \cite{force1,force2,harris,flow} The hysteresis loop was identified in the stretching and releasing processes of these experiments, suggesting that the long chain DNA undertakes different pathways for the above two processes. Nevertheless, in these experiments, the conformational transition was induced by external perturbations. Hence, the experiment to elucidate the solvent-induced conformational transition is greatly needed. Such an experiment provides insights into the thermodynamics and kinetics of the DNA coil-to-globule transition, and the molecular machinery of life. \cite{cpl99}

Despite the prediction from simulations \cite{sim1}, to our best knowledge, the direct measurement for the forward and reverse transitions between coiled and globular states has not been reported. In this work, we intend to elucidate the conformational hysteresis of a giant DNA mediated by condensing agents via integrating experimental and theoretical methods. In the experiment, PEG will be introduced as condensing agent, and both titration and backtitration will be conducted to investigate the folding and unfolding processes of DNA. Meanwhile, a simplified model will be employed to study the folding and unfolding processes through dynamics Monte Carlo simulation suggested by Milchev et al. \cite{rand} The density-of-state Monte Carlo method \cite{w-l} will be further applied to determine the stable conformational state in the hysteresis loop.

\section{Experiment}

Single chain observation on individual giant DNA molecules by use of fluorescence microscopy has been performed following the procedures as described in our previous reports. \cite{exp1,exp2} This method renders an opportunity to provide precise measurement for the chain conformation of free DNA molecules in the bulk solution. Thus, in this section, we will mainly describe the experimental method exploited to investigate the hysteresis on the conformational transition of DNA molecules.

In the experiment, the titration and backtitration of DNA solutions with PEG (Polyethylene glycol) are carried out to investigate the chain behavior of a giant DNA molecule in the presence of condensing agents. The experimental details are summarized as follows. A bacteriophage T4DNA (166 kbp, Nippon Gene) solution of 0.2 $\mu$ M (in base units), is obtained by dissolving the DNA in 300 mM NaCl (Nacalai Tesque) containing 0.1 $\mu$ M of the fluorescent dye 4', 6-diamidino-2-phenylindole (DAPI, Wako Chemical Industries). A series of PEG (with molecular weight 20000 Daltons from Kishida Chemical, Osaka) solutions are prepared in the range between 33 and 65 mg/mL. The experimental temperature is kept at $21^\circ $C. In the forward titration, the DNA solution is mixed with PEG of various concentrations, along with incubation for 60 or 1000 minutes. In the backtitration, the DNA solution is mixed with a PEG stock solution of 60 mg/mL first followed by incubation for 60 minutes to collapse all DNA molecules into compact state. In the next step, the compact DNA solution is added to deionized water in order to dilute the DNA-PEG mixture, and each diluted mixture is under incubation for 60 or 1000 minutes. 
The single molecular fluorescence images are utilized to count the number of coiled and compact DNA molecules in solution.

\section{Model and Monte Carlo Simulations}\label{model}

To abstract the essential mechanism on the observed hysteresis, Monte Carlo simulations are resorted. In the simulations, an isolated FENE (Finite Extension Nonlinear Elastic) chain (with eight monomers and thirty-two monomers) is considered, consisting of bond vibrational energy, repulsive L-J (Lennard-Jones) potential, and bending energy. \cite{stevens} The bending energy adjusts the chain stiffness, which reads $0.5 \kappa (\theta-\theta_0)$ where $\kappa$ is the strength of bending energy; $\theta$ is the bond angle; $\theta_0 = \pi$. The parameters of these potentials are obtained from reference \cite{stevens}. To mimic the poor solvent condition arising from condensing agents, an effective attractive interaction between any two monomers is incorporated, which takes the form of Yukawa potential, given by $V_{att}=-\epsilon \frac{\exp{(-Br)}}{r}$ where $\epsilon$ is the strength of attractive interaction; $B^{-1}$ is the characteristic length of the attractive interaction. In the calculations, the L-J diameter, the Boltzmann constant and temperature are assumed to be unity, and $B=1$. 

Monte Carlo simulations are carried out by using the traditional Metropolis and density-of-states (DOS) methods. In the Metropolis Monte Carlo simulations, the chain conformation is relaxed by using random walk for monomers, equivalent to the Rouse motion of a polymer chain as suggested by Milchev et al \cite{rand}, and a total of $1 \times 10^8$ moves for $N=8$ and $2 \times 10^8$ moves for $N=32$ are carried out for each case. In the DOS Monte Carlo method, we follow the algorithm suggested by Wang and Landau to compute the density of states $\Omega(E)$ \cite{w-l}, and the simulation is mainly conducted for shorter chains that give better statistics. 

\section{Results and discussion}

The conformational behavior of a giant DNA in the bulk solution is characterized by using single molecular images. Figure \ref{fig1} displays the fluorescence images of a single T4DNA for two distinct conformations (elongated chain on left and compact chain on right) in (a), the quasi-3D representation corresponding to the light intensity distribution in the fluorescence image in (b), and the schematic of the long-axis length, $L$, which is defined as the longest distance in the outline of DNA images  along with the blurred effect, c.a. 0.3 $\mu$ m, in (c). The conformations in figure \ref{fig1}(a) are two typical chain morphologies for coiled and globular states, respectively, with very different volumes. The structure of each conformation is further verified through the quasi-3D representation in figure \ref{fig1}(b). From these measurements, the chain size is determined based on the long-axis length $L$, which serves as an order parameter in the analysis of chain size distribution.

In relation to them, it has been confirmed in the previous studies that the bimodal distribution around the transition region is well characterized by the measurement of long-axis length, as well as the other measurements such as hydrodynamic radius, $R_h$, and radius of gyration, $R_g$. \cite{rgh} It is to be noted that we have performed fluorescence microscopy observation on the DNA molecules existing exclusively in the bulk solution. The low-resolution images are unavoidable for such a measurement. This is because the Brownian motion of the DNA in the bulk solution causes the blurring effect on the images. Nevertheless, the blurring effect (c.a., 0.3 mm) is smaller than the chain size (long DNA chains used in our experiment), and the measured conformation of single DNA molecules yields the precise information on the conformational transition. In contrast, the DNA absorbed onto the glass plate exhibits a much better resolution than those in bulk solutions. Through image resolutions, the chain molecules under Brownian motion in the bulk solution and those adsorbed onto the surface can be distinguished. Thus, in this manuscript, only the data of the DNA molecules in the bulk solution are shown.

To examine the conformational behavior along the forward titration and backtiration processes, the chain size distribution is investigated. Figure \ref{fig2} plots the histograms of the long-axis length $L$ of T4DNA molecules with an increase of PEG concentration during forward titration from top to bottom (45 mg/mL, 48 mg/mL, 52 mg/mL and 60 mg/mL) on left column, and with a decrease of PEG concentration during backtitration from bottom to top (51 mg/mL, 48 mg/mL, and 45 mg/mL) on right column after incubation of 60 minutes for each concentration. In forward titration, the chain size distribution for low PEG concentrations is basically single modal, and the peak is located at a larger $L$ ($\approx 1.5$ $\mu$ m).  As PEG concentration is increased to $48$ mg/mL, the bimodal distribution starts to emerge.  When PEG concentration is increased further to 52 mg/mL, the bimodal distribution become more pronounced, suggesting that DNA molecules fold and undergo a discontinuous transition with coexistent coiled and globular conformational states. At a high PEG concentration (60 mg/mL), the distribution is single modal, and the peak position shifts toward a smaller $L$, c.a. $0.5 \mu$ m for mean $L$. The result indicates that all DNA molecules are transformed from coiled state at a low PEG concentration (45 mg/mL) to globular state at a high PEG concentration (60 mg/mL) through the transition as seen in the first-order phase transition. 

In backtitration, the qualitative features are reversed, in which more DNA molecules unfold as the PEG concentration is decreased. Nevertheless, the quantitative difference between folding (titration) and unfolding (backtitration) processes can be discerned for [PEG] = 45 and 48 mg/mL after 60-min incubation, for example. In these PEG concentrations, the profiles of long-axis $L$ distribution of forward titration (folding) is significantly different from those of backtitration (unfolding). The statistical weight of globular state remains higher for the backtitration (unfolding) process.

In addition to the amount of condensing agent, it is noted that the statistical distribution of coiled and globular states is also dependent on the processes of titration and/or incubation time. In the following, the investigation is conducted during forward titration and backtitration processes. For each process, one PEG concentration is selected to study the effect of incubation time. Figure \ref{fig3} displays the histogram of DNA chain size at [PEG] = 50 mg/mL during the forward titration process (on left) and at [PEG] = 51 mg/mL during the backtitration process (on right) for different incubation time, 60 (top) and 1000 (bottom) minutes. In the forward titration process, the probability of finding globular conformations is low for the 60-min incubation time. After increasing the incubation time to 1000 min, the bimodal profile of the chain size distribution becomes more pronounced with an increase of globular chains in solution. On contrary, in the case of backtitration, the measured histogram is not sensitive to incubation time. The results may be attributed to different conformational behavior in the titration and backtitration processes.

The above analysis based on histograms allows us to estimate the mean statistical weight of the two coexistent conformational states during the forward titration and backtitration processes. Figure \ref{fig4} displays the fraction of elongated (or coiled) state $f_{coil}$ obtained from forward titration and backtitration, denoted by solid and open symbols, respectively, after incubation for 60 minutes ($\circ$) and 1000 minutes ($\Delta$); lines provide as a visual guide. In the forward titration, $f_{coil}$ decreases as PEG concentration is increased (facilitating chain folding) for a given incubation time. Furthermore, $f_{coil}$ decreases with an increase of incubation time from 60 to 1000 minutes, and the titration curve shifts to smaller $f_{coil}$. Unlike forward titration, the result of backtitration exhibits different behavior. As PEG concentration is decreased in the backtitration, fewer DNA chains tend to elongate; consequently, the number of elongated chains is smaller compared to the forward titration at a given PEG concentration. It is noticeable that for the unfolding process, $f_{coil}$ is dependent on the PEG concentration, but becomes less sensitive to incubation time. The conformational hysteresis in the titration loop indicates different pathways involved in the folding and unfolding processes of a giant DNA molecule in the presence of condensing agents. Note some DNA molecules remain the compact form in the dilute PEG solution even for the longest incubation time under the unfolding process.  

To elucidate the conformational hysteresis, the simplified semiflexible chain model introduced in section \ref{model} is studied. The mean chain size is first calculated through the random-walk Monte Carlo method in the following way. In the folding process, the simulation begins with the coil state, and then the monomer-monomer attraction $\epsilon$ is increased gradually till the chain collapses into compact state. After the chain collapses totally, the reverse process is applied to unfold the chain by decreasing $\epsilon$, and $\epsilon$ is decreased till $\epsilon=0$. Figure \ref{fig5} plots the variation of mean squared radius of gyration, divided by a factor $\Gamma$, with $\epsilon$ and some snapshots for $\kappa=2$ ($\Gamma=1$) and $10$ ($\Gamma=10$) when $N=8$, and for $\kappa=20$ ($\Gamma=100$) when $N=32$, as marked,  for the folding (solid symbols) and unfolding (open symbols) processes; lines provide as a visual guide. The factor $\Gamma$ is introduced to collapse all the curves in the same plot. As the monomer-monomer attraction is increased, the chain starts to contract. For a more flexible chain (e.g., $\kappa=2$ and $N=8$), the transition between an elongated and a compact state is reversible. As the chain stiffness is increased (e.g., $\kappa=10$ and $N=8$, or $\kappa=20$ and $N=32$), a conformational hysteresis occurs, indicating distinct folding and unfolding pathways for both processes. Besides, we find that once the chain collapses, unfolding the chain (particularly for longer chains, e.g. $N=32$) back to its elongated state becomes difficult. These results are consistent with the experimental observations in figure \ref{fig4}.  

The simulation snapshots exhibit the chain structures during the folding and unfolding processes. In figure \ref{fig5}, for the less stiff chain, such as $\kappa=2$ and $N=8$, the folding and unfolding processes are reversible, and the chain tends to form cyclic conformation after it collapses, e.g., $\epsilon=8$. A further increase of $\epsilon$ forces the chain to contract more, e.g. $\epsilon=26$. For a stiffer chain, such as $\kappa=10$ and $N=8$, the polymer collapses from an elongated chain (e.g., $\epsilon = 8$) to a compact ring structure (e.g., $\epsilon=42$) in the folding process. During the unfolding process, the compact chain takes a more expanded ring structure (e.g., $\epsilon=22$) before it becomes totally elongated. For a longer and stiffer chain, e.g., $\kappa=20$ and $N=32$, the elongated chain conformation persists till $\epsilon \approx 56$, and the collapsed chain adopts a highly compact structure consisting of four smaller rings intercalating each other, as shown in $\epsilon = 61$. In the unfolding process, the chain makes a transition from the compact structure (e.g., $\epsilon = 47$ similar to that $\epsilon = 61$), to two rings (e.g., $\epsilon = 11$) before it is totally elongated. These results again show the distinct pathways for folding and unfolding processes.

A further calculation is conducted to identify the stable conformational state in the hysteresis loop. Figure \ref{fig6} compares the energy distribution (divided by a factor $\gamma$) of the elongated and compact states obtained from the Metropolis method (solid lines and $\gamma=1$) with that of the DOS method (dotted line and $\gamma=2$) for $\kappa=10$ and $\epsilon=30$ when $N=8$. The result shows that the DOS method, independent of the choice of initial conformations, agrees well with the energy distribution of the compact state in the hysteresis, suggesting that the compact state is thermodynamically more stable and the elongated state is a metastable state. The folding process is impeded during the length of our simulations, probably, due to the energy trapping for a polymer chain.

From the simulations, the effect of chain stiffness on the density of states is examined. Figure \ref{fig7} plots the renormalized density of states $\Omega(E)$ for $\kappa=5$ and $10$, as marked, and for $\epsilon = 0$ (solid lines) and $4$ (dotted lines) when $N=8$; the straight line denotes the Boltzmann factor. The crossing point between the line of the Boltzmann factor and each curve reflects the rough peak position of the energy distribution function for a given set of $\kappa$ and $\epsilon$. In contrast to a more flexible chain, the energy distribution of a stiffer chain appears at a higher energy, and the value of its density of states is smaller at around the line equivalent to Boltzmann factor. As the monomer-monomer attraction is increased, the energy distribution shifts to a lower energy, but the density of states of a stiff chain remains smaller than that of a flexible chain. 

The different kinetic pathways for the folding and unfolding processes of a giant DNA have been addressed extensively in literature. \cite{kin1,kin2,kin3} The nucleation is required for the first step of the folding process \cite{landau} as in the intramolecular nucleation involved in polymer crystallization \cite{muthu,frekel}, but not for the unfolding process. An increase of the attraction between monomers stabilizes the nucleation site and facilitates the subsequent growth of the folded structure. In this work, a complementary explanation from thermodynamic standpoint is given as follows. A decrease of the density of states by increasing chain stiffness is corresponding to attenuation of possible chain conformations in a semiflexible chain. We speculate that as the total number of conformations becomes low, those conformations relevant to the folding pathway may also decrease simultaneously. As a result, the folding process is impeded, in particular, if the folding related conformations bear higher energies. With an increase of monomer-monomer attraction, the density of states becomes larger, which increases the number of possible conformations as well as the probability of folding. Furthermore, the low density of states of a semiflexible chain sheds new light on its ordered compact structure. For a flexible chain polymer, the chain molecule collapses into a liquid-like globule, whereas for a semiflexible chain, it forms a highly-ordered toroid. \cite{dna1,dna3} Such a highly ordered structure in the compact globule state may link with the low density-of-states in a semiflexible DNA molecule.

\section{Conclusions}

The titration and backtitration with PEG (Polyethylene glycol) have been conducted to investigate the conformational behavior of a giant DNA molecule mediated by condensing agents in the bulk solution. A pronounced conformational hysteresis is observed for folding and unfolding processes, by increasing and decreasing PEG concentration, respectively, in experiment. To abstract the essential mechanism on the observed hysteresis, a semiflexible chain model is studied by using the random-walk Monte Carlo simulation to mimic the Brownian motion of monomers. The semiflexible chain molecule is modeled as a non-linear elastic spring subjected to a bending energy. A monomer-monomer attractive interaction is incorporated to model poor solvent conditions. The simulation is conducted in the following way. The folding process is proceeded as the monomer-monomer attraction is increased, and after the chain is collapsed into the compact state, the unfolding process is followed by decreasing the monomer-monomer attraction. As chain becomes stiff enough, a conformational hysteresis occurs in the random-walk Monte Carlo simulations. The density-of-states Monte Carlo method shows that the compact state is thermodynamically more stable than the elongated one in the hysteresis loop. Our results suggest that the chain stiffness induced conformational hysteresis may reduce of the chain conformations relevant to the folding pathway of a semiflexible chain. As a result, the folding process becomes thermodynamically unfavorable because of the low probability to reach the intermediate state.

\section{Acknowledgments}

CYS received partial support for this work from the City University of New York PSC-CUNY grants \# 67666-0036 and \# 68544-00-37, NSF Garcia MRSEC at SUNY Stony Brook, and NY START grant for the Center of Engineered Polymeric Materials at CSI and Institute of Macromolecular Assembly.
This work was supported by Japan Society for the Promotion of Science (JSPS) under a Grant-in-Aid for Creative Scientific Research (Project No. 18GS0421).

\newpage
\begin{figure}
\caption {Fluorescence images of a single T4DNA existing in the bulk solution with two distinct conformations in (a) (elongated chain on left and compact chain on right), the quasi-3D representation corresponding to the light intensity distribution in the fluorescence image in (b), and the schematic of the long-axis length, L, which is defined as the longest distance in the outline of DNA images in (c).}
\label{fig1}

\caption {Plot of the histograms of the long-axis length $L$ of T4DNA molecules with an increase of PEG concentration during forward titration from top to bottom (45 mg/mL, 48 mg/mL, 52 mg/mL and 60 mg/mL) on left column, and with a decrease of PEG concentration during backtitration from bottom to top (51 mg/mL, 48 mg/mL, and 45 mg/mL) on right column after incubation of 60 minutes for each concentration. All of the data were obtained from the DNA molecules existing in the bulk solution.
}
\label{fig2}

\caption {Plots of the histograms of DNA chain size in the bulk solution at [PEG] = 50 mg/mL during the titration process (on left) and at [PEG] = 51 mg/mL during the backtitration process (on right) for different incubation time, 60 (top) and 1000 (bottom) minutes.}
\label{fig3}

\caption {Variation of the fraction of elongated chains $f_{coil}$ in the bulk solution with PEG concentration obtained from (forward) titration and backtitration, denoted by solid and open symbols, respectively, after incubation for 60 minutes ($\circ$) and 1000 minutes ($\Delta$) (lines provide as a visual guide.)}

\label{fig4}

\caption {Variation of mean squared radius of gyration, divided by a factor $\Gamma$, with $\epsilon$ and some snapshots for $\kappa=2$ ($\Gamma=1$) and $10$ ($\Gamma=10$) when $N=8$, and for $\kappa=20$ ($\Gamma=100$) when $N=32$, as marked, under the forward (solid symbols) and backward (open symbols) processes; lines provide as a visual guide, and the factor $\Gamma$ is introduced to collapse all the curves in the same plot.}
\label{fig5}

\caption {Comparison of the energy distribution (divided by a factor $\gamma$) of the elongated and compact states obtained from the Metropolis method (solid lines; $\gamma=1$) with that of the DOS method (dotted line; $\gamma=2$) for $\kappa=10$ and $\epsilon=30$ when $N=8$; $\gamma$ is introduced to differentiate the results from the two different methods.}
\label{fig6}
\end{figure}

 \begin{figure}
\caption {Plot of the renormalized density of states for $\kappa=5$ and $10$, as marked, and for $\epsilon = 0$ (solid lines) and $4$ (dotted lines) when $N=8$; the straight line denotes the Boltzmann factor.}
\label{fig7}
\end{figure}

\end{document}